# Conditioning a Recurrent Neural Network to synthesize musical instrument transients

**Lonce Wyse**
Communication and New Media
National University of Singapore
`lonce.wyse@nus.edu.sg`

**Muhammad Huzaifah**
NUS Graduate School of Integrative
Sciences and Engineering
National University of Singapore
`e0029863@u.nus.edu`

## ABSTRACT

A Recurrent Neural Network (RNN) is trained to predict sound samples based on audio input augmented by control parameter information for pitch, volume, and instrument identification. During the generative phase following training, audio input is taken from the output of the previous time step, and the parameters are externally controlled allowing the network to be played as a musical instrument. Building on an architecture developed in previous work, we focus on the learning and synthesis of transients – the temporal response of the network during the short time (tens of milliseconds) following the onset and offset of a control signal. We find that the network learns the particular transient characteristics of two different synthetic instruments, and furthermore shows some ability to interpolate between the characteristics of the instruments used in training in response to novel parameter settings. We also study the behavior of the units in hidden layers of the RNN using various visualization techniques and find a variety of volume-specific response characteristics.

## 1. INTRODUCTION

When musical wind instrument sounds are initiated by blowing air through or across a mouthpiece, the time it takes for the system to reach a stable resonant state is referred to as an "attack" transient. When energy ceases to be put in to the system, the time it takes for the instrument to return to rest is a "decay" transient. The attack is typically complex with different frequency components reaching their steady states via different amplitude trajectories ("envelopes"). The attack characteristics vary significantly across the way the instrument is articulated with tongue and breath, and across different instruments. The attack characteristics are an important perceptual indicator used by listeners to identify playing style and the instrument being played (Grey [1]).

Transients are interesting from a modeling perspective in part because of their non-instantaneous response to the articulation parameters. There is no simple mapping from breath pressure to the sound signal the instrument radiates, but there is rather a state-dependent temporal evolution of the sound that follows sudden changes in the parameters.

In our previous work, (Wyse [2]), we developed a recurrent neural network (RNN) for modeling musical instrument sound generation. RNN's were chosen because they are structured and often used to model sequences such as digital sound samples. The training was conditioned on pitch, volume, and instrument ID in addition to the audio stream so that during generation, the parameters could be used to control synthesis. However, in this previous work, only steady-state tones were used during training, so none of the specific instrument transient characteristics were learned.

In this paper, we train the network on two different synthetic instruments that, in addition to having different harmonic structures, also have distinct attack and decay times that follow sudden changes in the control parameter that we use as a proxy for breath pressure, and that we refer to herein as "volume". We also explore the activations of hidden units in response to parameter changes during generation and find interesting patterns such as volume-specific responses.

## 2. MODELING

### 2.1 Architecture

The architecture is the same as that used in Wyse [2] and is summarized here (Figure 1). It is a stacked RNN composed of 4 layers of Gated Recurrent Units (GRU) (Cho et al. [3]) sandwiched between dense layers after the input and before the output.

The input is a vector of 4 components at each time step (sample rate=16000 Hz) representing the audio sample, the pitch, volume, and instrument ID. Audio is mu-law encoded with 256 values and normalized to [0,1], pitch consisted of the 13 chromatic notes spanning the octave between E4 (fundamental frequency=~330 Hz) to E5 (fundamental frequency=~660 Hz) and normalized to floating point values in [0,1], and volume was mapped exponentially from a 40 dB range to [0,1]. The output of the network is a vector of length 256, where each component represents a mu-law encoded sample value. For training, the audio target signal was coded one-hot, and for generation, the maximally valued output was taken as the audio sample.

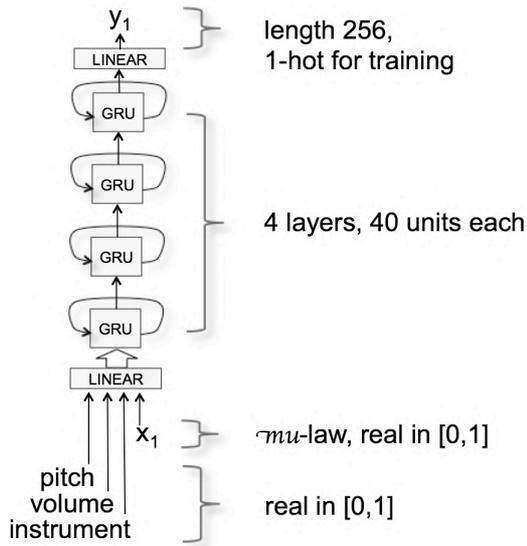

**Figure 1**. The network consists of 4 layers of 40 GRU units each. A four-dimensional vector is passed through a linear layer as input and the output is a one-hot encoded audio sample.

## 2.2 Training Data

Training data consists of 2 instruments, each with 13 different pitches and 25 different volume levels. The two instruments each have their own harmonic structure and transient duration. "SynthEven" was constructed of even harmonics only and "SynthOdd" with odd harmonics only. The waveforms can be seen in Figure 2.

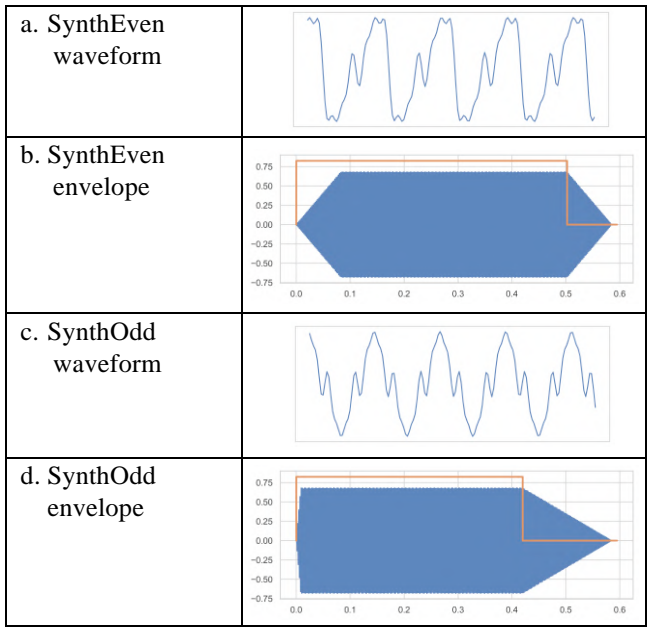

**Figure 2**. Characteristics of the two synthetic instruments used to train the neural network. The synthetic instruments have different wave forms (a,c) and different attack and decay slopes (b,d). The transients form in correspondence to a sudden change in the volume parameter (orange line).

In addition, SynthEven is constructed with attack and decay transients with a linear slope of ±10 volume units/sec, while SynthOdd has attacks with slope +100 and a decay with slope -5 volume units/sec. (Figure 2b, d). Note that transients all have constant slope which means that the duration of the transients depends on the steady-state volume of the tones.

## 3. RESULTS

### 3.1 Steady-state volume

The focus of the current work is on the transient responses to sudden changes in the volume parameter. However, for musical playability, we still require the trained instruments to track the volume parameter at steady state as well as over smooth changes across its range. Figure 3 shows the output of the network for the two instruments in response to various input volume parameter patterns.

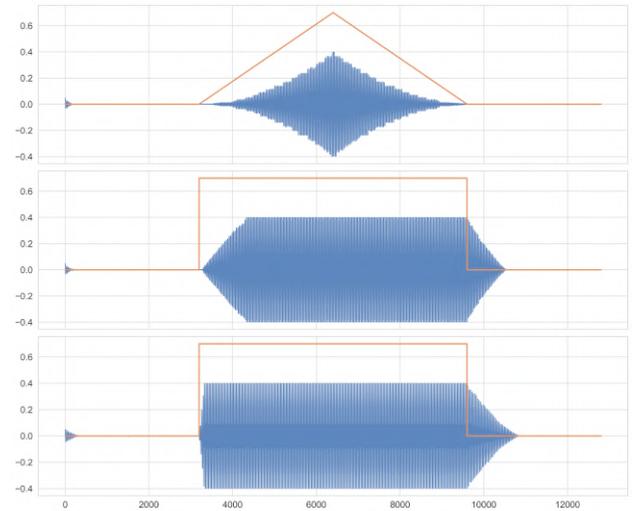

**Figure 3**. From top to bottom: a) SynthEven, pitch=0.5 (B♭4, ~466 Hz), response to smooth volume change over 400 ms. b) SynthEven, pitch=0.5, response to sudden volume parameter changes c) SynthOdd, pitch=0.5, response to sudden volume parameter changes. In all three scenarios the volume parameter (orange line) was adjusted between a minimum of 0 and a maximum of 0.7. The x-axis depicts sample number.

The network is thus capable of learning to respond to changes in the conditioning input with state-dependent response extended in time. Furthermore, since the network is trained on two different instruments with different transient characteristics, the temporal response depends on a second conditioning parameter specifying the instrument.

## 4. HIDDEN LAYER PATTERNS

Next, we take a closer look at the hidden unit activation patterns during synthesis to understand the network computations. With only 40 nodes per layer, we can visualize the entire network activation patterns over time.

Visualizations show that almost all nodes oscillate with individual characteristic waveforms as illustrated in Figure 4. That waveform is similar across different conditioning parameters, except that the periodicity of the waveforms tracks the period of the pitch specified by the pitch parameter. The node activations show almost none of the frequency selectivity we find in hair cells and neurons along the hearing pathways in animal and human brains. This was somewhat unexpected as distributions of frequency specific patterns have been found in other types of learning networks that operate on audio data and learn efficient representations (e.g. Lewicki [4]; Hoshen et al. [5]; Sailor and Patil [6]). In this paper, we just note the pitch-locked oscillatory pattern, but focus on responses to volume input.

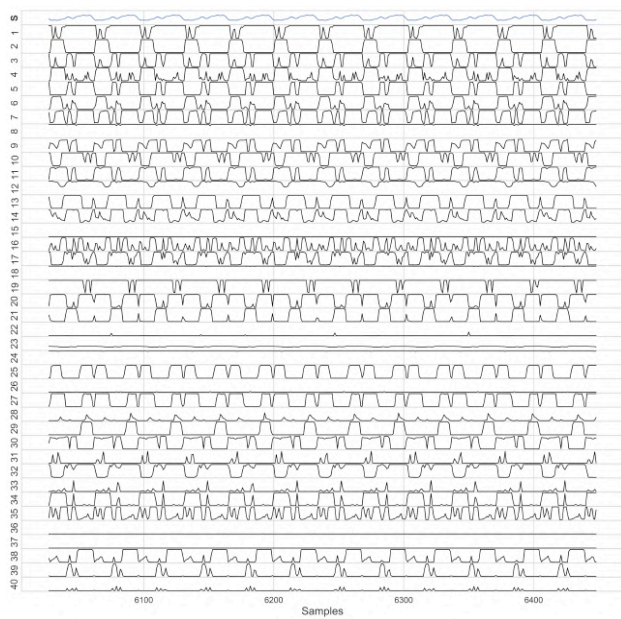

**Figure 4**. The 40 hidden unit responses of a section of generated audio labelled **S** spanning approximately 400 samples. Each hidden unit displays a characteristic response waveform, the shape of which changes in response to volume level and instrument. In contrast, varying the pitch parameter changes the period of the hidden waveform without altering its overall pattern.

The four hidden layers in the network show distinctly different patterns in response to volume changes. Layer 1 (the layer closest to the input) consists of units most of which oscillate with amplitudes correlated with the volume parameter. A few have a "DC offset" (some positive, some negative) that also tracks the volume parameter. However, none show volume-specific selectivity. Instead their responses merely track volume.

Each succeeding deeper layer shows more complex structure. At the last hidden layer (prior to the linear layer connected to the output units), patterns of volume selectivity are clearly visible (Figure 6).

Considering only the amplitude of oscillation and not the DC offset, we see that node 37, for example, is maximally responsive to low volume; nodes 39 and 40 are responsive to mid-range volumes only, although node 39 has an upwardly shifted and wider range sensitivity than node 40. Node 15 and 23 only respond to high volumes. We have found that these volume response sensitivities are the same whether the volume is ramping up or ramping down. We have seen no responses sensitive to the direction of slowly changing volume. These patterns are also robust across pitch.

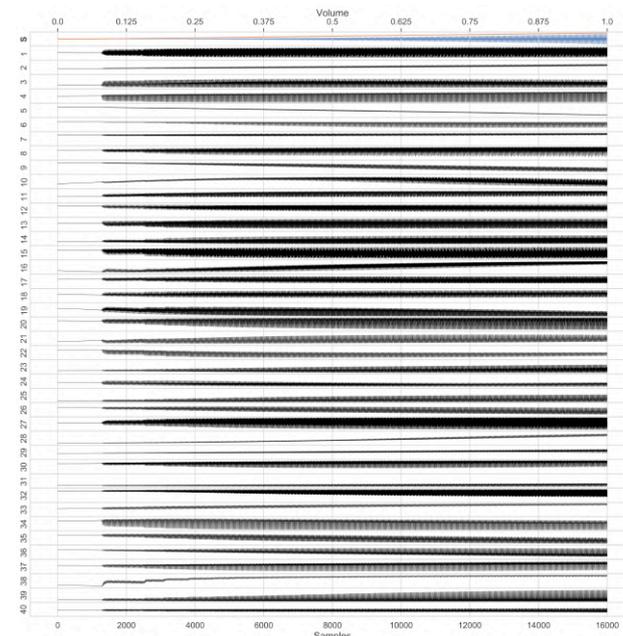

**Figure 5.** Layer 1 (shallowest) hidden unit responses to a continuous increase in volume. All nodes oscillate with amplitude that correlates with the input parameter (as well as with the amplitude of the output signal).

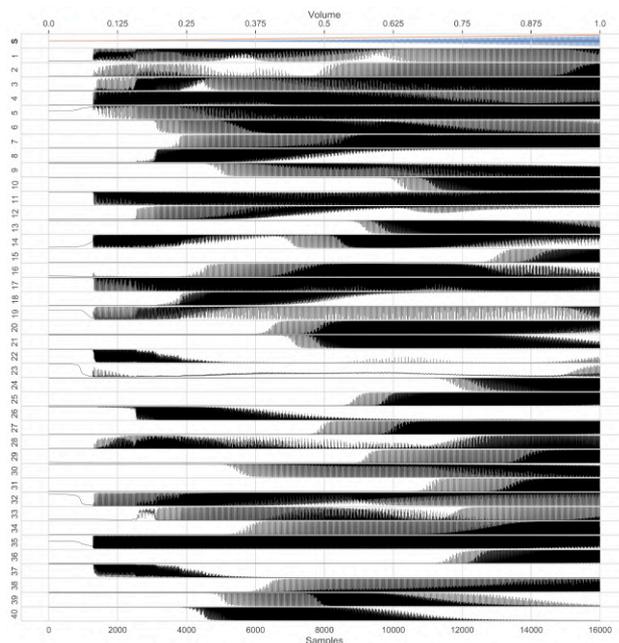

**Figure 6.** Layer 4 (deepest) hidden unit responses to a continuous increase in volume. Individual hidden units clearly show unique volume selectivity.

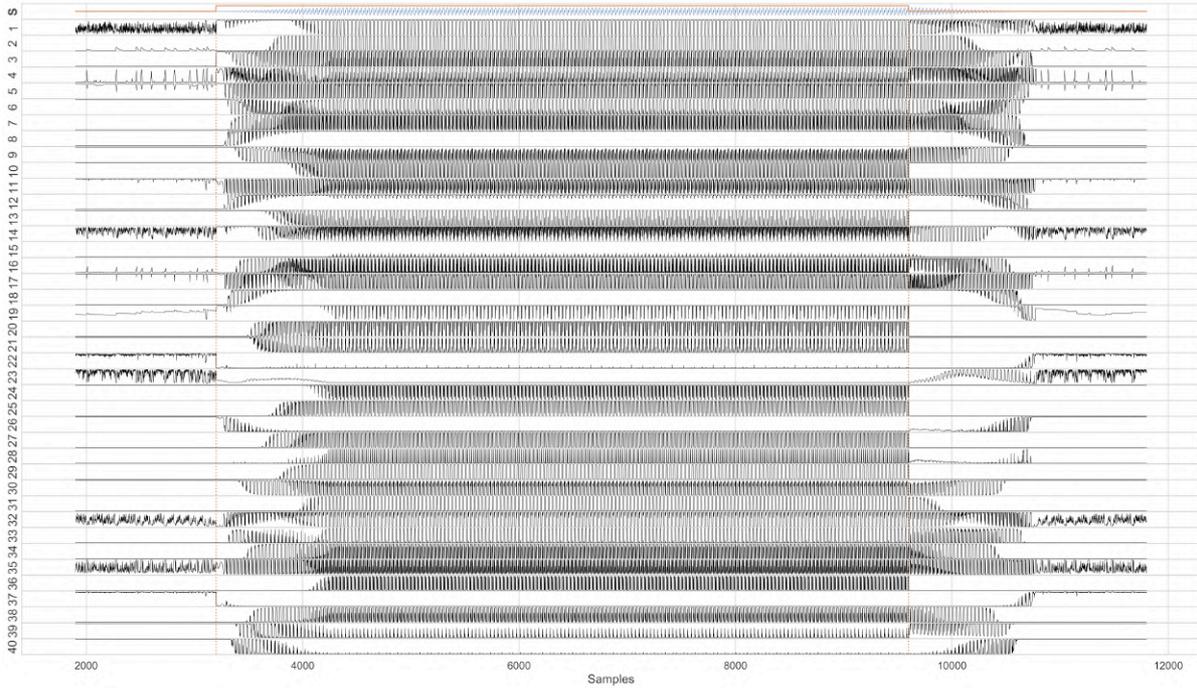

**Figure 7.** The response of units in the deepest hidden layer to the sudden onset and offset of the volume parameter demarcated by the dotted orange lines, with the synthesized signal and time evolution of the volume parameter shown in the top row. Conditional parameters used were as follows: instID=SynthEven, pitch=0.5, volume=0 to 0.8 to 0.

### 4.1 Transient responses to abrupt volume changes

The responses of hidden neurons to abrupt changes in the volume parameter are more complex, as would be expected, since during the attack and decay transients there is a "mismatch" between the volume parameter and the volume of the output signal. The mismatch is negligible during steady state or the slow sweeping volume changes considered above.

Figure 7 shows the deepest hidden layer for the response of SynthEven (i.e. trained with symmetrical attack and decay slopes) to a sudden onset and offset of the volume parameter. The output signal **S** (along the top of the figure) is close to the signal used to train this parameter pattern, although further tests showed the overall shape and length of the decay being somewhat inconsistent and fairly dependent on the parameter combination and the priming signal (a single random sample) used to initialize the synthesis process. Note that the transients in the signal output amplitude are an order of magnitude faster than the volume sweep used in Figure 5 and Figure 6.

One notable feature of this map is that although the output signal amplitude is roughly symmetric following the onset and offset of the volume parameter, the response of the units in this layer are not. Far fewer nodes show an immediate change in activation following the onset of the volume parameter than do to the offset of the parameter. Such behavior is reflected in Figure 7 where following the offset of the volume, certain units immediately cease to oscillate (e.g. 10, 13, 20, 25), while others continue to respond with the decaying amplitude. Similar divergent patterns were not apparent with a volume onset.

Some of the patterns of volume selectivity that we found during the slow volume sweep (Figure 6) are still visible during transients in the same units. For example, unit 40 responds to low volume in the output signal during both the attack and decay transients, just as it did during the sweep. Units 9 and 10 maintain their relative volume selectivity during the attack transient as for the sweep. However, unit 10 shuts off immediately with the volume parameter, while unit 9 is active until the output signal almost disappears.

In general, the attack portion of the signal was more reliably generated and consistent in timing than the decay portion. This might be due to the asymmetry of the training regime. During training, the volume always went from 0 to the target volume level for the attack, and from the target volume level to 0 prior to the decay. This means that for the portion of the signal following the step change, the attacks were exposed to 25 different volume levels, while the decay portion occurred while the volume was a 0 for all examples, no matter what the steady-state volume before the decay was. A better training scheme might be to train on step ups from non-zero values for attacks, and more importantly, to train on smaller steps down (not all the way to zero) for decays.

## 5. CONCLUSIONS

Wyse [2] developed an RNN that learns to generate signals for different synthetic and natural instruments (Engel et al. [7]) conditioning on pitch and volume so that after training, the models could be played under controls similar to musical instruments. We showed that training examples could be sparse in pitch, and trained only on steady state signals, yet the model responded quickly and accurately to pitch parameter values and sequences it had never seen during training. In the current work, we have shown that the same model can also capture attack and decay transients where the response to the conditioning input is extended over time.

Transients are proving more difficult to capture in this small model. They are less robust, more sensitive to priming signals (used to initialize the hidden state) and noise, and do not seem to generalize as easily as pitch or steady-state volume to parameter values and sequences not see during training. Future work will be necessary to increase the robustness of these results and to apply the network to natural musical instrument data.

We also explored the activation patterns of nodes in hidden layers in response to volume changes and found more selectivity in response to volume levels than was apparent for pitch, gaining a deeper understanding of how the network computes its sound-modeling task, which will help guide the further development of this type of network for learning interactive musical sound synthesis models.

**Acknowledgments**

This research was supported a Singapore MOE Tier 2 grant, "Learning Generative and Parameterized Interactive Sequence Models with Recurrent Neural Networks," by an NVIDIA Corporation Academic Programs GPU grant.